\documentclass[12pt,twoside]{article}
\usepackage[english]{babel}
\usepackage{graphicx,rotating,epsfig}
\usepackage{a4p}
\usepackage{xspace}

\def\ra{\rightarrow} 
\def\GeV  {\ensuremath{\mathrm{ Ge\kern -0.1em V } }}
\def\GeVc2{\ensuremath{\mathrm{ Ge\kern -0.1em V }\kern -0.2em /c^2 }}
\def\MeVc2{\ensuremath{\mathrm{ Me\kern -0.1em V }\kern -0.2em /c^2 }}
\newcommand{\MT}{\ensuremath{M_{\mathrm{t}}}}
\newcommand{\MTll}{\ensuremath{\MT^{\mathrm{\ell\ell}}}}
\newcommand{\MTlj}{\ensuremath{\MT^{\mathrm{\ell\mbox{+}jets}}}}
\newcommand{\MTjj}{\ensuremath{\MT^{\mathrm{alljets}}}}
\newcommand{\MTmet}{\ensuremath{\MT^{\mathrm{MEt}}}}

\newcommand{\fb}{\ensuremath{\mathrm{fb}^{-1}}}
\newcommand{\ttbar}{\ensuremath{t\overline{t}}}
\newcommand{\WbWb}{\ensuremath{W^+ b W^- \overline{b}}}
\newcommand{\ljt}{\ensuremath{\ell\nu b q q^{\prime} \overline{b}}}
\newcommand{\had}{\ensuremath{q q^{\prime} b q q^{\prime} \overline{b}}}
\newcommand{\dil}{\ensuremath{\ell^{+}\nu b\ell^{-}\overline{\nu}\overline{b}}}
\newcommand{\ttljt}{\ensuremath{\ttbar\ra\WbWb\ra\ljt}}
\newcommand{\ttdil}{\ensuremath{\ttbar\ra\WbWb\ra\dil}}
\newcommand{\tthad}{\ensuremath{\ttbar\ra\WbWb\ra\had}}

\newcommand{\RunI}{\hbox{Run\,I}}
\newcommand{\RunII}{\hbox{Run\,II}}
\newcommand{\measStatSyst}[3]{\ensuremath{#1 \pm #2\thinspace(\textrm{stat}) \pm #3\thinspace(\textrm{syst})}\xspace}
\newcommand{\gevcc}[1]  {\ensuremath{#1~\mathrm{GeV}/c^{2}}}
\newcommand{\et}     {\ensuremath{E_{T}}\xspace}
\newcommand{\met}    {\mbox{$\protect \raisebox{.3ex}{$\not$}\et$}\xspace}

\newcommand{\pte}    {\ensuremath{p_{T}^{\mathrm{lep}}}}
\newcommand{\Lxy} {L_{XY}}
\begin{document}

\begin{center}
  {\LARGE FERMI NATIONAL ACCELERATOR LABORATORY}
\end{center}

\begin{flushright}
       FERMILAB-CONF-13-164-PPD-TD \\  
       TEVEWWG/top 2013/01 \\
       CDF Note 10976\\
          D\O\ Note 6381\\
       \vspace*{0.05in}
         August 2013 \\ 
\end{flushright}

\vskip 1cm

\begin{center}
  {\LARGE\bf 
    Combination of CDF and D\O\ results 
    on the mass of the top quark using up to \boldmath{$8.7\:\fb$} at
    the Tevatron\\
  }
  \vfill
  {\Large
    The Tevatron Electroweak Working Group\footnote{The Tevatron Electroweak 
    Working Group can be contacted at tev-ewwg@fnal.gov.\\  
    \hspace*{0.20in} More information can
    be found at {\tt http://tevewwg.fnal.gov}.} \\
    for the CDF and D\O\ Collaborations\\
  }
\end{center}
\vfill
\begin{abstract}
\noindent
  We summarize the current top-quark mass measurements from the CDF and
  D\O\ experiments at Fermilab.  We combine published
  \RunI\ (1992--1996) measurements with the most precise published and preliminary
  \RunII\ (2001--2011) measurements based on data sets corresponding to up to $8.7~\fb$ of $p\bar{p}$ collisions.
 Taking correlations of uncertainties into account, and
 combining the statistical and systematic uncertainties, the resulting
  preliminary Tevatron average mass of the top quark is $\MT =
  \gevcc{173.20 \pm 0.87}$, corresponding to a relative precision of $0.50\%$.
  
\end{abstract}

\vfill




\section{Introduction}
\label{sec:intro}

This note reports the Tevatron average top-quark mass obtained by
combining the most precise published and preliminary measurements of
the top-quark mass. It is an update of the combination presented in
Ref.~\cite{TeVTopComboPRD}, where further details can be found.
The ATLAS and CMS collaborations have also performed a combination of their most recent
top quark mass measurements~\cite{lhccombi}.

The CDF and D\O\ collaborations have performed several direct experimental measurements of the
top-quark mass (\MT) using data collected at the Tevatron
proton-antiproton collider located at the Fermi National Accelerator
Laboratory.  These pioneering measurements were first based on approximately
$0.1~\fb$ of \RunI\ data
~\cite{Mtop1-CDF-di-l-PRLa}-\cite{Mtop1-D0-allh-PRL}
collected from 1992 to 1996,
and included results from the decay channels \tthad\ (alljets),
\ttljt\ ($\ell$+jets), and 
\ttdil\ ($\ell\ell$), where $\ell=e$ or $\mu$.  
Decays with $\tau \to e, \mu$ are included in the direct $W \to e$ and
$W \to \mu$ channels. 
In \RunII\ (2001--2011), many top mass measurements have been performed, and
those considered here are the most
recent results in these channels, using up to $8.7~\fb$ of data for CDF
(corresponding to the full CDF Run II dataset)
~\cite{
Mtop2-CDF-MEt-new,
Mtop2-CDF-l+jt-pub-new,
Mtop2-CDF-di-l-pub,
Mtop2-CDF-allh-pub},
and up to $5.4~\fb$ of data for D\O\
~\cite{ 
Mtop2-D0-l+ja-final,
Mtop2-D0-l+jt-new2,
Mtop2-D0-di-l}.
The CDF analysis based upon charged particle tracking for exploiting
the transverse decay length 
of $b$-tagged jets ($L_{XY}$) and the transverse momentum of
electrons and muons from $W$ boson decays ($\pte$) uses
a data set corresponding to a luminosity of
1.9~$\fb$~\cite{Mtop2-CDF-trk}, and
there are no plans to update this analysis.
The D\O\ \RunII\  measurements presented in this note include the most
recent Run II measurement in  
the $\ell\ell$~\cite{Mtop2-D0-di-l}  channel using 5.4~fb$^{-1}$ of data 
and in the $\ell$+jets channel \cite{Mtop2-D0-l+jt-new2} 
with 3.6~fb$^{-1}$ of data. Both results are now published.
Since the combination performed in 2011~\cite{Mtop-tevewwgSum11}, a new
final state signature was introduced by CDF that requires events to
possess missing transverse energy ($\met$) and jets, but no identified
lepton (``MEt'')~\cite{Mtop2-CDF-MEt-new,Mtop2-CDF-MEt-published}.  
This sample is statistically independent from the previous three CDF
channels.

With respect to the July 2011 combination\,\cite{Mtop-tevewwgSum11} and the published version of the 
combination~\cite{TeVTopComboPRD}, the \RunII\ CDF measurement in the 
$\ell$+jets channel has been updated using 8.7~fb$^{-1}$ 
of data, an improved analysis technique, and improved jet energy resolution \cite{Mtop2-CDF-l+jt-pub-new}. 
The CDF measurement in the MEt channel was updated to use the full
\RunII\ data set for CDF of  8.7~fb$^{-1}$ of data as 
well~\cite{Mtop2-CDF-MEt-new}.
The now published \RunII\ CDF measurements in the $\ell\ell$ channel\,\cite{Mtop2-CDF-di-l-pub} and alljets 
channel\,\cite{Mtop2-CDF-allh-pub}  are unchanged. The measurement 
based on charged particle tracking~\cite{Mtop2-CDF-trk} was
incorporated as described in the past 
combinations~\cite{Mtop-tevewwgSum11}. 
From the corresponding analysis only
the measurement of the top quark mass using the mean decay length $\Lxy$ of
$B$ hadrons in $b$-tagged lepton+jets events has been used. It
is independent of energy information in the calorimeter, and its main
source of systematic uncertainty is uncorrelated with the dominant
ones from the jet energy scale calibration in other measurements. This
measurement of $m_t$ is essentially uncorrelated with the higher
precision CDF result from the lepton+jets channel. The overlap between
the data samples used for the decay-length method and the lepton+jets
sample has therefore no effect. 

The Tevatron average top-quark mass is obtained by combining five published
\RunI\ measurements~\cite{Mtop1-CDF-di-l-PRLb, Mtop1-CDF-di-l-PRLb-E,
  Mtop1-D0-di-l-PRD, Mtop1-CDF-l+jt-PRD, Mtop1-D0-l+jt-new1,
  Mtop1-CDF-allh-PRL} with four published \RunII\ CDF 
results~\cite{Mtop2-CDF-l+jt-pub-new,Mtop2-CDF-di-l-pub,Mtop2-CDF-allh-pub, 
Mtop2-CDF-trk}, one preliminary \RunII\ CDF
result~\cite{Mtop2-CDF-MEt-new}, and two published 
\RunII\ D\O\ results~\cite{Mtop2-D0-l+jt-new2,
Mtop2-D0-di-l}.
This combination 
supersedes previous
combinations~\cite{Mtop-tevewwgSum11,Mtop1-tevewwg04,Mtop-tevewwgSum05,
  Mtop-tevewwgWin06,Mtop-tevewwgSum06, Mtop-tevewwgWin07, Mtop-tevewwgWin08, 
  Mtop-tevewwgSum08, Mtop-tevewwgWin09,Mtop-tevewwgSum10}. 

The definition and evaluation of the systematic uncertainties and the understanding of the
correlations among channels, experiments, and Tevatron runs is the outcome of many years of 
joint work between the CDF and D\O\ collaborations and is described in detail
elsewhere~\cite{TeVTopComboPRD}.

The input measurements and uncertainty categories used in the combination are 
detailed in Sections~\ref{sec:inputs} and~\ref{sec:uncertainty}, respectively. 
The correlations assumed in the combination are discussed in 
Section~\ref{sec:corltns} and the resulting Tevatron average top-quark mass 
is given in Section~\ref{sec:results}.  A summary is presented
in Section~\ref{sec:summary}.
 
\section{Input Measurements}
\label{sec:inputs}

Twelve measurements of \MT\ used in this combination are shown in Table~\ref{tab:inputs}.
The \RunI\ measurements all have relatively large statistical
uncertainties and their systematic uncertainties are dominated by the
total jet energy scale (JES) uncertainty.  In \RunII\, both CDF and
D\O\ take advantage of the larger \ttbar\ samples available and employ
new analysis techniques to reduce both of these uncertainties.  In
particular, the \RunII\ D\O\ analysis in the $\ell$+jets channel and the 
\RunII\ CDF analyses in the $\ell$+jets, alljets, and MEt channels 
constrain the response of light-quark jets using the kinematic information from $W\ra
qq^{\prime}$ decays (so-called {\em in situ}
calibration)~\cite{Mtop1-CDF-l+jt-PRD,Abazov:2006bd}. Residual JES uncertainties associated with
$p_{T}$ and $\eta$ dependencies as well as uncertainties specific to
the response of $b$ jets are treated separately. The
\RunII\ D\O\ $\ell\ell$ measurement uses the JES determined in the
$\ell$+jets channel by {\em in situ} calibration~\cite{Mtop2-D0-di-l}.

\vspace*{0.10in}

\begin{table}[t]
\caption[Input measurements]{Summary of the measurements used to determine the
  Tevatron average \MT.  Integrated luminosity ($\int \mathcal{L}\;dt$) has units of
  \fb, and all other numbers are in $\GeVc2$.  The uncertainty categories and 
  their correlations are described in Section~\ref{sec:uncertainty}.  The total systematic uncertainty 
  and the total uncertainty are obtained by adding the relevant contributions 
  in quadrature. ``n/a'' stands for ``not applicable'', ``n/e'' for ``not evaluated''.}
\label{tab:inputs}
\begin{center}

\renewcommand{\arraystretch}{1.30}
{\tiny
\begin{tabular}{l|ccc|cc|cccc|cc|c} 
\hline \hline
       & \multicolumn{5}{c|}{{\RunI} published} 
       & \multicolumn{6}{c|}{{\RunII} published} 
       & \multicolumn{1}{c}{{\RunII} prel.}  \\ 
       & \multicolumn{3}{c|}{ CDF } 
       & \multicolumn{2}{c}{ D\O\ }
       & \multicolumn{4}{|c|}{ CDF }
       & \multicolumn{2}{c|}{ D\O\ }
       & \multicolumn{1}{c}{ CDF }
        \\

                      &     $\ell$+jets   & $\ell\ell$     &   alljets  &     $\ell$+jets &   $\ell\ell$  &    $\ell$+jets     &     $\ell\ell$  &    alljets  &     Lxy     &   $\ell$+jets      &   $\ell\ell$    &      MEt \\
\hline                                                                                                                                                       
$\int \mathcal{L}\;dt$&       0.1  &     0.1  &    0.1  &     0.1  &     0.1 &      8.7    &       5.6 &     5.8  &         1.9 &         3.6 &       5.3 &      8.7   \\
\hline                                                                                                                                                       
Result                &   176.1    & 167.4    &186.0    & 180.1    & 168.4   &      172.85 &    170.28 &  172.47  &      166.90 &      174.94 &    174.00 &    173.95\\
\hline                                                                                                                                                       
\shortstack{ {\em In situ} light-jet cali- \\
bration (iJES)}                   &        n/a &      n/a & n/a     &      n/a &      n/a &       0.49 &      n/a  &     0.95 &         n/a &        0.53 &      0.55  &      1.05  \\
\shortstack{ Response to $b$/$q$/$g$ \\
jets (aJES)}                          &        n/a &      n/a & n/a     &      0.0 &      0.0 &       0.09 &      0.14 &     0.03 &         n/a &        0.0  &      0.40 &      0.10 \\
\shortstack{ Model for $b$ jets  \\
(bJES)}                               &        0.6 &      0.8 & 0.6     &      0.7 &      0.7 &       0.16 &      0.33 &     0.15 &         n/a &        0.07 &      0.20 &      0.17  \\
 \shortstack{ Out-of-cone correction \\
     (cJES)}                          &        2.7 &      2.6 & 3.0     &      2.0 &      2.0 &       0.21 &      2.13 &     0.24 &        0.36 &        n/a  &      n/a  &      0.18  \\
\shortstack{ Light-jet response (2) \\ 
    (dJES)}                               &        0.7 &      0.6 & 0.3     &      2.5 &      1.1 &       0.07 &      0.58 &     0.04 &        0.06 &        0.63 &      0.56 &      0.04  \\
\shortstack{ Light-jet response (1) \\
 (rJES)}                               &        3.4 &      2.7 & 4.0     &      n/a &      n/a &       0.48 &      2.01 &     0.38 &        0.24 &        n/a  &      n/a  &      0.40  \\
\shortstack{ Lepton modeling  \\
(LepPt)}                              &        n/e &      n/e & n/e     &      n/e &      n/e &       0.03 &      0.27 &     n/a    &        n/a  &        0.17 &      0.35 &      n/a   \\
\shortstack{ Signal modeling \\
 (Signal)}                             &        2.6 &      2.9 & 2.0     &      1.1 &      1.8 &       0.61 &      0.73 &     0.62 &        0.90 &        0.77 &      0.86 &      0.64\\
 \shortstack{ Jet modeling \\
  (DetMod)}                             &        0.0 &      0.0 & 0.0     &      0.0 &      0.0 &       0.0  &      0.0  &     0.0  &        0.0  &        0.36 &      0.50 &      0.0 \\
\shortstack{ Offset \hspace{0.5cm} \\
 (UN/MI)}                              &        n/a &      n/a &    n/a  &      1.3 &      1.3 &       n/a  &      n/a  &     n/a  &        n/a  &        n/a  &      n/a  &      n/a  \\
\shortstack{ Background from \\
theory (BGMC)}                        &        1.3 &      0.3 &    1.7  &      1.0 &      1.1 &       0.12 &      0.24 &     0.0  &        0.80 &       0.18  &      0.0  &      0.0   \\
 \shortstack{ Background based on\\
data (BGData)}                     &        0.0 &      0.0 &    0.0  &      0.0 &      0.0 &      0.16  &      0.14 &     0.56 &        0.20 &       0.23  &      0.20 &      0.12  \\
 \shortstack{ Calibration method\\
(Method)}                      &        0.0 &      0.7 & 0.6     &      0.6 &      1.1 &       0.00 &      0.12 &     0.38&         2.50 &        0.16 &      0.51 &      0.31  \\
 \shortstack{ Multiple interactions\\
 model (MHI)}                          &        n/e &      n/e &    n/e  &      n/e &      n/e &       0.07 &      0.23 &     0.08&         0.0  &        0.05 &      0.0  &      0.18  \\
\hline                                                                                                                                                       
 \shortstack{ Systematic uncertainty\\
 (Syst)}                   &        5.3 &      4.9 & 5.7    &       3.9 &      3.6 &       0.98 &      3.09 &     1.49 &        2.90 &        1.24 &      1.44 &      1.35 \\
\shortstack{ Statistical uncertainty\\
 (Stat)}                   &        5.1 &     10.3 &10.0    &       3.6 &     12.3 &       0.52 &      1.95 &     1.43 &        9.00 &        0.83 &      2.36 &      1.26 \\
\hline                                                                                                                                                       
Total uncertainty                     &        7.3 &     11.4 &11.5    &        5.3 &    12.8 &       1.11 &      3.79 &     2.06 &        9.46 &       1.50 &      2.76 &      1.85\\ 
\hline
\hline
\end{tabular}
}
\end{center}
\end{table}

The D\O\ Run~II $\ell$+jets analysis uses the JES determined from the
external calibration derived from $\gamma$+jets events as an
additional Gaussian constraint to the {\em in situ} calibration. Therefore,
the total resulting JES uncertainty is split into one part obtained from the 
{\em in situ} calibration and another part determined from the external calibration.
To do this, the measurement without external
JES constraint has been combined iteratively with a pseudo-measurement
using the method of Refs.~\cite{Lyons:1988, Valassi:2003} 
that uses only the external calibration in a way that the combination gives the total JES uncertainty. 
The splitting obtained in this way is used to assess both the statistical part of the JES uncertainty and the part of 
the JES uncertainty due to the external calibration constraint~\cite{Mtop2-D0-comb}.

The $\Lxy$ technique developed by CDF 
uses the decay length of $B$ mesons from $b$-tagged jets.
While the statistical sensitivity of this analysis is not as good as
that of the more
traditional methods, this technique has the advantage that it is almost entirely independent of
JES uncertainties since it
uses primarily tracking information.
\vspace*{0.10in}

The D\O\ \RunII\ $\ell$+jets result is a combination of the 
published Run~IIa (2002--2005) measurement ~\cite{Mtop2-D0-l+ja-final} with 1~fb$^{-1}$ 
of data and the result obtained with 2.6~fb$^{-1}$ of data from Run~IIb (2006--2007)~\cite{Mtop2-D0-l+jt-new2}.
This analysis includes an additional particle response correction on top of the
standard {\it in situ} calibration.
The D\O\ \RunII\ $\ell\ell$ result is based on a neutrino weighting technique using 5.4~fb$^{-1}$
of \RunII\  data~\cite{Mtop2-D0-di-l}. 
\vspace*{0.10in}

Table~\ref{tab:inputs} lists the individual uncertainties of each result,
subdivided into the categories described in the next Section.  The
correlations between the inputs are described in
Section~\ref{sec:corltns}.


\section{Uncertainty Categories}
\label{sec:uncertainty}

We employ uncertainty categories similar to what was used for the previous Tevatron
average~\cite{TeVTopComboPRD,Mtop-tevewwgSum11}, with small
modifications to better account for their correlations.
They are divided such that sources of systematic uncertainty that share the same or similar origin are 
combined as explained in Ref.~\cite{TeVTopComboPRD}. 
For example, the {\it Signal modeling} ({\it Signal}) category
discussed below includes the uncertainties from  different systematic
sources that are correlated due to their origin in the modeling of the simulated signal samples.

 Some systematic uncertainties have been separated into multiple
categories to accommodate specific types of correlations.
For example, the jet energy scale (JES) uncertainty is subdivided
into six components to more accurately accommodate our
best understanding of the relevant correlations between input measurements. 

For this note we use the new systematic naming scheme described in Ref.~\cite{TeVTopComboPRD}. In parentheses, the 
old names of the  systematic uncertainties are provided.  There is a one-to-one matching between the new and old systematic 
definitions of categories.

\vspace*{0.10in}

\begin{description}
  \item[Statistical uncertainty (Statistics):] The statistical uncertainty associated with the
    \MT\ determination.
 \item[{\em In situ} light-jet calibration (iJES):] That part of the
   JES uncertainty that originates from
   {\em in situ} calibration procedures and is uncorrelated among the
   measurements.  In the combination reported here, it corresponds to
   the statistical uncertainty associated with the JES determination
   using the $W\ra qq^{\prime}$ invariant mass in the CDF \RunII\
   $\ell$+jets, alljets, and  MEt measurements and the D\O\ Run~II
   $\ell\ell$ and $\ell$+jets
   measurements. 
   For the D\O\ Run~II $\ell$+jets measurement, it also includes the uncertainty
   coming from the MC/data difference in jet response that is uncorrelated
   with the other D\O\ Run~II measurements. 
   Residual JES uncertainties arising from effects
   not considered in the {\em in situ} calibration are included in other
   categories. 
  \item[Response to \boldmath{$b/q/g$} jets (aJES):] That part of the JES
    uncertainty that originates from 
    average differences in detector electromagnetic over hadronic ($e/h$)
    response for hadrons produced in the fragmentation of $b$-jets and light-quark 
    jets. 
  \item[Model for \boldmath{$b$} jets (bJES):] That part of the JES uncertainty that originates from
    uncertainties specific to the modeling of $b$ jets and that is correlated
    across all measurements.  For both CDF and D\O\ this includes uncertainties 
    arising from 
    variations in the semileptonic branching fractions, $b$-fragmentation 
    modeling, and differences in the color flow between $b$-quark jets and light-quark
    jets.  These were determined from \RunII\ studies but back-propagated
    to the \RunI\ measurements, whose {\it Light-jet response (1)}  uncertainties ({\it rJES}, see below) were 
    then corrected to keep the total JES uncertainty constant.
  \item[Out-of-cone correction (cJES):] That part of the JES uncertainty that originates from
    modeling uncertainties correlated across all measurements.  
    It specifically includes the modeling uncertainties associated with light-quark 
    fragmentation and out-of-cone corrections. For D\O\ \RunII\ measurements,
    it is included in the {\it Light-jet response (2) (dJES)} category.
  \item[Light-jet response (1) (rJES):] The remaining part of the JES
    uncertainty that covers the absolute calibration for CDF's \RunI\
    and \RunII\ measurements. It also includes small contributions
    from the uncertainties associated with modeling multiple
    interactions within a single bunch crossing and corrections for
    the underlying event. 
%
  \item[Light-jet response (2) (dJES):] That part of the JES
    uncertainty that includes D\O's \RunI\ and \RunII\ calibrations of
    absolute response (energy dependent), the relative response
    ($\eta$-dependent), and the out-of-cone showering correction that 
    is a detector effect. This uncertainty term for CDF includes only
    the small relative response calibration ($\eta$-dependent) for
    \RunI\ and \RunII. 
%
  \item[Lepton modeling (LepPt):] The systematic uncertainty arising from uncertainties
    in the scale of lepton transverse momentum measurements. It was not
    considered as a source of systematic uncertainty in the \RunI\
    measurements. 
  \item[Signal modeling (Signal):] The systematic uncertainty arising from uncertainties
    in \ttbar\ modeling that is correlated across all
    measurements. This includes uncertainties from variations of the amount of initial and 
    final state radiation and from the choice of parton density function used
    to generate the \ttbar\ Monte Carlo samples
    that calibrate each method. For D\O, it also includes the uncertainty 
    from higher-order corrections evaluated from a comparison of \ttbar\ samples generated by {\textsc MC@NLO} ~\cite{MCNLO} and 
    {\textsc ALPGEN}~\cite{ALPGEN}, both interfaced to {\textsc HERWIG}~\cite{HERWIG5,HERWIG6} for the simulation of parton showers and hadronization.
    In this combination, the systematic uncertainty arising from a variation of the 
  phenomenological description of color reconnection (CR) between final state  particles \cite{CR,Skands:2009zm} 
  is included in the {\it Signal modeling} category.
  The CR uncertainty is obtained by taking the difference between the {\textsc PYTHIA}\,6.4 tune ``Apro" and the {\textsc PYTHIA}\,6.4 tune 
``ACRpro" that differ only in the CR model. 
This uncertainty was not evaluated in Run~I since the Monte Carlo
generators available at that time did not allow for variations of the CR model.
These measurements therefore do not include this source of systematic uncertainty. Finally, the systematic 
uncertainty associated with variations of the MC generator used to calibrate the mass extraction method is added. 
It includes variations observed when 
    substituting {\textsc PYTHIA} 
    \cite{PYTHIA4,PYTHIA5,PYTHIA6} 
    (\RunI\ and \RunII) 
    or {\textsc ISAJET}~\cite{ISAJET} (\RunI) for {\textsc HERWIG}~\cite{HERWIG5,HERWIG6} when 
    modeling the \ttbar\ signal.  
%
  \item[Jet modeling (DetMod):] The systematic uncertainty arising from uncertainties 
in the modeling of jet interactions in the detector in the MC
simulation. For D\O\, this includes uncertainties from jet resolution
and identification. 
Applying jet algorithms to MC events, CDF finds that the resulting
efficiencies and resolutions closely match those in data. The small
differences propagated to $\MT$ lead to a negligible uncertainty of
0.005~GeV, which is then ignored.  

 \item[Background based on data (BGData):] This includes 
    uncertainties associated with the modeling using data of the QCD
    multijet background in the alljets,
    MEt, and $\ell$+jets channels and
    the Drell-Yan background in the $\ell\ell$ channel. 
    This part is uncorrelated between experiments.

  \item[Background from theory (BGMC):] 
This systematic uncertainty on the background originating from theory
(MC) takes into account the    
 uncertainty in modeling the background sources. It is correlated between
    all measurements in the same channel, and includes uncertainties on the background composition, normalization, and shape of different components, e.g., the 
uncertainties from the modeling of the $W$+jets background in the $\ell$+jets channel  
associated with variations of the factorization scale used to simulate $W$+jets events. 

  \item[Calibration method (Method):] The systematic uncertainty arising from any source specific
    to a particular fit method, including the finite Monte Carlo statistics 
    available to calibrate each method. 
  \item[Offset (UN/MI):] This uncertainty is specific to D\O\ and includes the uncertainty
    arising from uranium noise in the D\O\ calorimeter and from the
    multiple interaction corrections to the JES.  For D\O\ \RunI\ these
    uncertainties were sizable, while for \RunII, owing to the shorter
    calorimeter electronics integration time and {\em in situ} JES calibration, these uncertainties
    are negligible.
  \item[Multiple interactions model (MHI):] The systematic uncertainty arising from a mismodeling of 
  the distribution of the number of collisions per Tevatron bunch crossing owing to the 
  steady increase in the collider instantaneous luminosity during data-taking. 
  This uncertainty has been separated from other sources to account for the fact that 
  it is uncorrelated between experiments.

\end{description}
These categories represent the current preliminary understanding of the
various sources of uncertainty and their correlations.  We expect these to 
evolve as we continue to probe each method's sensitivity to the various 
systematic sources with improving precision.

\section{Correlations}
\label{sec:corltns}

The following correlations are used for the combination:
\begin{itemize}
  \item The uncertainties in the {\it Statistical uncertainty (Stat)} and
    {\it Calibration method (Method)}
    categories are taken to be uncorrelated among the measurements.
  \item The uncertainties in the {\it In situ light-jet  
    calibration (iJES)}
    category are taken to be uncorrelated among the measurements
    except for D0's $\ell\ell$ and $\ell$+jets measurements, where
    this uncertainty is taken to be 100\% correlated since the
    $\ell\ell$ measurement uses the JES calibration  determined in
    $\ell$+jets channel.  
  \item The uncertainties in the {\it Response to $b$/$q$/$g$ jets (aJES)}, {\it Light-jet response (2) (dJES)}, {\it Lepton modeling (LepPt)},
    and {\it Multiple interactions model (MHI)} categories are taken
    to be 100\% correlated among all \RunI\ and all \RunII\ measurements 
    within the same experiment, but uncorrelated between \RunI\ and \RunII\
    and uncorrelated between the experiments.
  \item The uncertainties in the {\it Light-jet response (1) (rJES)}, {\it Jet modeling (DetMod)}, and {\it Offset (UN/MI)} categories are taken
    to be 100\% correlated among all measurements within the same experiment 
    but uncorrelated between the experiments.
  \item The uncertainties in the {\it Backgrounds estimated from theory (BGMC)} category are taken to be
    100\% correlated among all measurements in the same channel.
  \item The uncertainties in the {\it Backgrounds estimated from data (BGData)} category are taken to be
    100\% correlated among all measurements in the same channel and same run period, but uncorrelated between the experiments.
  \item The uncertainties in the {\it Model for $b$ jets (bJES)}, {\it Out-of-cone correction (cJES)}, and {\it Signal modeling (Signal)}
    categories are taken to be 100\% correlated among all measurements.
\end{itemize}
Using the inputs from Table~\ref{tab:inputs} and the correlations specified
here, the resulting matrix of total correlation coefficients is given in
Table~\ref{tab:coeff}.

\begin{table}[t]
\caption[Global correlations between input measurements]{The matrix of correlation coefficients used to determine the
  Tevatron average top-quark mass.}
\begin{center}
\renewcommand{\arraystretch}{1.30}
\tiny
\begin{tabular}{l|ccc|cc|cccc|cc|c}
\hline \hline
     & \multicolumn{5}{c}{{\RunI} published} 
         & \multicolumn{6}{|c|}{{\RunII} published} 
         & \multicolumn{1}{c}{{\RunII} preliminary}   \\
      & \multicolumn{3}{c|}{ CDF } 
         & \multicolumn{2}{c}{ D\O\ }
         & \multicolumn{4}{|c|}{ CDF } 
	 & \multicolumn{2}{c|}{ D\O\ }
         & \multicolumn{1}{c}{ CDF } 
         \\

                &  $\ell$+jets    &    $\ell\ell$ &   alljets &    $\ell$+jets  &   $\ell\ell$   &   $\ell$+jets  &   $\ell\ell$   & alljets   &     $\Lxy$   &  $\ell$+jets      & $\ell\ell$  &  MEt \\
                                                                                                                  
\hline                                                                                                            
CDF-I $\ell$+jets    &   1.00  &    0.29  &    0.32  &    0.26  &    0.11  &    0.49  &    0.54  &    0.25  &    0.07  &    0.21  &    0.12  &    0.27 \\
CDF-I $\ell\ell$     &   0.29  &    1.00  &    0.19  &    0.15  &    0.08  &    0.29  &    0.32  &    0.15  &    0.04  &    0.13  &    0.08  &    0.17 \\
CDF-I alljets        &   0.32  &    0.19  &    1.00  &    0.14  &    0.07  &    0.30  &    0.38  &    0.15  &    0.04  &    0.09  &    0.06  &    0.16 \\
D\O-I $\ell$+jets    &   0.26  &    0.15  &    0.14  &    1.00  &    0.16  &    0.22  &    0.27  &    0.12  &    0.05  &    0.14  &    0.07  &    0.12 \\
D\O-I $\ell\ell$     &   0.11  &    0.08  &    0.07  &    0.16  &    1.00  &    0.11  &    0.13  &    0.07  &    0.02  &    0.07  &    0.05  &    0.07 \\
CDF-II $\ell$+jets   &   0.49  &    0.29  &    0.30  &    0.22  &    0.11  &    1.00  &    0.48  &    0.29  &    0.08  &    0.30  &    0.18  &    0.33 \\
CDF-II $\ell\ell$    &   0.54  &    0.32  &    0.38  &    0.27  &    0.13  &    0.48  &    1.00  &    0.25  &    0.06  &    0.11  &    0.07  &    0.26 \\
CDF-II alljets       &   0.25  &    0.15  &    0.15  &    0.12  &    0.07  &    0.29  &    0.25  &    1.00  &    0.04  &    0.16  &    0.10  &    0.17 \\
CDF-II $\Lxy$     &   0.07  &    0.04  &    0.04  &    0.05  &    0.02  &    0.08  &    0.06  &    0.04  &    1.00  &    0.06  &    0.03  &    0.04 \\
D\O-II $\ell$+jets    &   0.21  &    0.13  &    0.09  &    0.14  &    0.07  &    0.30  &    0.11  &    0.16  &    0.06  &    1.00  &    0.39  &    0.18 \\
D\O-II $\ell\ell$    &   0.12  &    0.08  &    0.06  &    0.07  &    0.05  &    0.18  &    0.07  &    0.10  &    0.03  &    0.39  &    1.00  &    0.11 \\
CDF-II MEt     &   0.27  &    0.17  &    0.16  &    0.12  &    0.07  &    0.33  &    0.26  &    0.17  &    0.04  &    0.18  &    0.11  &    1.00 \\

\hline
\hline
\end{tabular}
\end{center}
\label{tab:coeff}
\end{table}

The measurements are combined using a program implementing two 
independent methods: 
a numerical $\chi^2$ minimization and 
the analytic best linear unbiased estimator (BLUE) method~\cite{Lyons:1988, Valassi:2003}. 
The two methods are mathematically equivalent.
It has been checked that they give identical results for
the combination. The BLUE method yields the decomposition of the uncertainty on the Tevatron $\MT$ average in 
terms of the uncertainty categories specified for the input measurements~\cite{Valassi:2003}.

\section{Results}
\label{sec:results}

The resultant combined value for the top-quark mass is
\begin{eqnarray}
\nonumber
\MT=\gevcc{\measStatSyst{173.20}{0.51}{0.71}}. 
\end{eqnarray}
Adding the statistical and systematic uncertainties
in quadrature yields a total uncertainty of $\gevcc{0.87}$, corresponding to a
relative precision of 0.50\% on the top-quark mass.
It has a $\chi^2$ of 8.5 for 11 degrees of freedom, corresponding to
a probability of 67\%, indicating good agreement among all input
measurements.  The breakdown of the uncertainties is 
shown in Table\,\ref{tab:BLUEuncert}.  The total statistical and systematic  
uncertainties are reduced relative to the 
Summer 2011 combination \cite{Mtop-tevewwgSum11} and the published combination~\cite{TeVTopComboPRD} 
due to the increase of the CDF data samples  in the $\ell$+jets and MEt analyses and better 
treatment of JES corrections in the $\ell$+jets analysis. 
 
\begin{figure}[p]
\begin{center}
\includegraphics[width=0.8\textwidth]{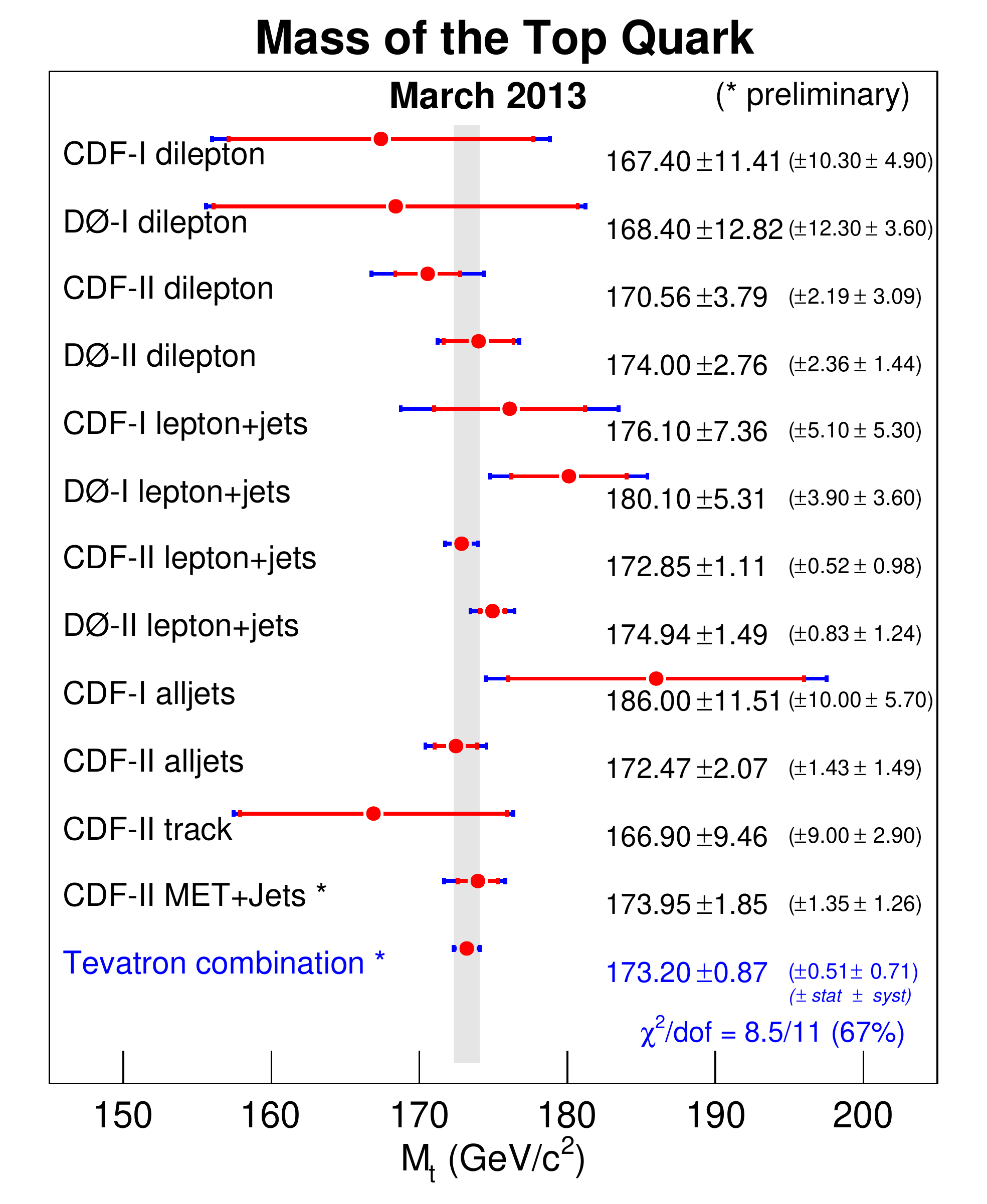}
\end{center}
\caption[Summary plot for the Tevatron average top-quark mass]
  {Summary of the input measurements and resulting Tevatron average
   mass of the top quark.}
\label{fig:summary} 
\end{figure}

The pull and weight for each of the inputs, as obtained from the
combination with the BLUE method, are listed in Table~\ref{tab:stat}.
The input measurements and the resulting Tevatron average mass of the top 
quark are summarized in Fig.~\ref{fig:summary}.
\vspace*{0.10in}

\begin{table}[tbh]
\caption{\label{tab:BLUEuncert} 
Summary of the Tevatron combined average $\MT$. The uncertainty categories are 
described in the text. The total systematic uncertainty and the total
uncertainty are obtained  
by adding the relevant contributions in quadrature.}
\begin{center}
\begin{tabular}{lc} \hline \hline
  & Tevatron combined values (GeV/$c^2$) \\ \hline
 $\MT$            & 173.20 \\ \hline
 {\em In situ} light-jet calibration (iJES)           &  0.36 \\
 Response to $b$/$q$/$g$ jets (aJES)                  &  0.09  \\
 Model for $b$ jets       (bJES)                  &  0.11 \\  
 Out-of-cone correction (cJES)                  &  0.01 \\ 
 Light-jet response (2) (dJES)                  &  0.15 \\ 
 Light-jet response (1) (rJES)                  &  0.16 \\
 Lepton modeling  (LepPt)                       &  0.05 \\
 Signal modeling  (Signal)                      &  0.52  \\
 Jet modeling (DetMod)                          &  0.08 \\
 Offset (UN/MI)                                 &  0.00 \\
 Background from theory (BGMC)                  &  0.06\\
 Background based on data (BGData)              &  0.13\\
 Calibration method (Method)                    &  0.06 \\
 Multiple interactions model (MHI)              &  0.07 \\ \hline
 Systematic uncertainty (syst)                  &  0.71 \\ 
 Statistical uncertainty (stat)                 &  0.51 \\ \hline
 Total  uncertainty                             &  0.87 \\
   \hline \hline
\end{tabular}
\end{center}
\end{table}

\begin{figure}
\begin{center}
\includegraphics[width=1.0\textwidth]{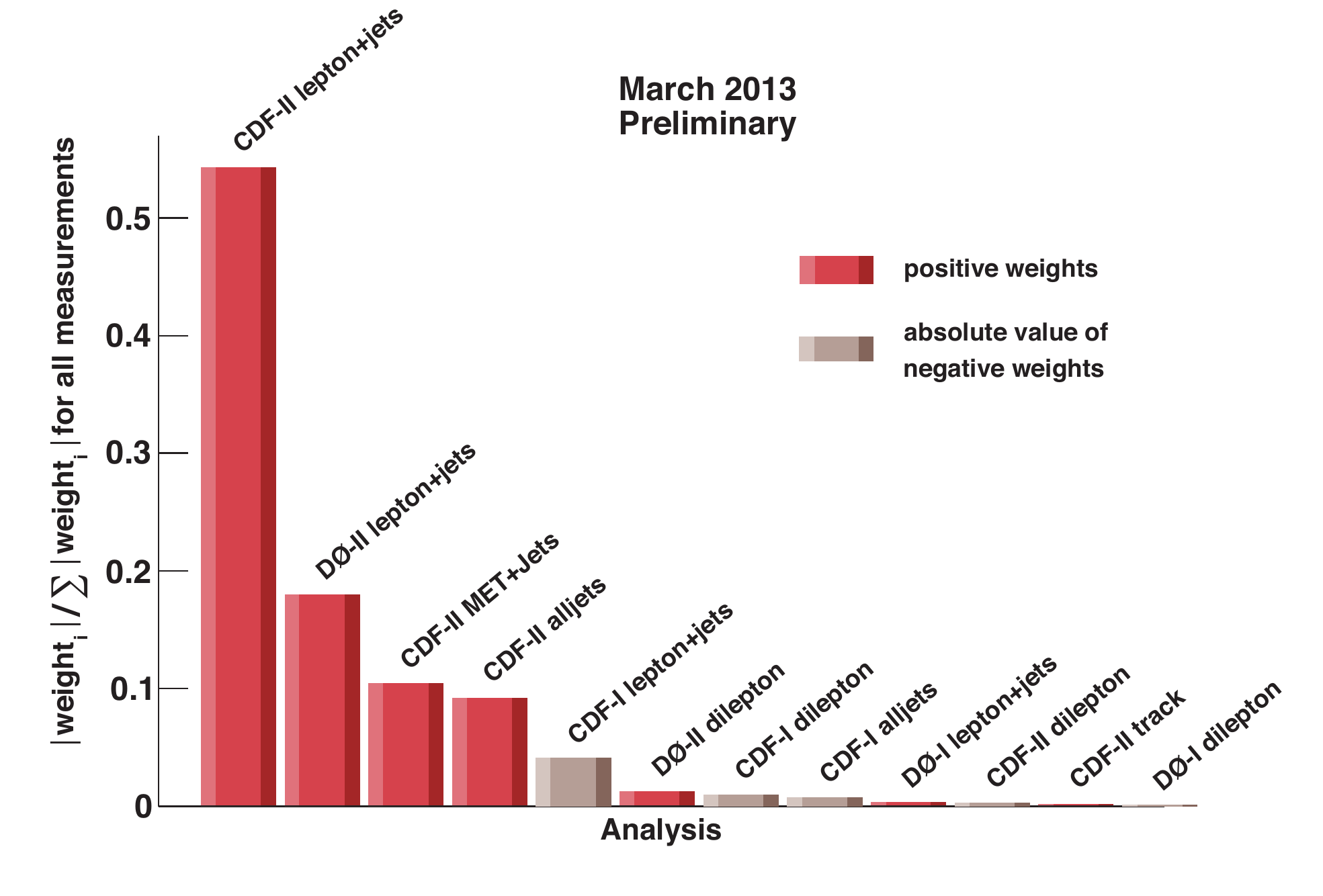}
\end{center}
\caption{Relative weights of the input measurements in the
  combination. The relative weights have been obtained by dividing the absolute value of each measurement weight by the sum over all measurements of the absolute values
of the weights. Negative weights are represented by their absolute
value, but using a grey color.}
\label{fig:Weights} 
\end{figure}

The weights of some of the measurements are negative, which occurs if
the correlation between two measurements 
is larger than the ratio of their total uncertainties.
In these instances the less precise measurement 
will acquire a negative weight.  While a weight of zero means that a
particular input is effectively ignored in the combination, channels
with a negative weight affect the resulting $\MT$ central value and help reduce the total
uncertainty~\cite{Lyons:1988}. 
To visualize the weight each measurement carries in the combination, Fig.\,\ref{fig:Weights} shows the
absolute values of the weight of each measurement divided by the sum of the absolute values of the weights
of all input measurements. Negative weights are represented by bins
with a different (grey) color. 
We note, that due to correlations between the uncertainties the relative weights
of the different input channels may be significantly different from
what one could expect from the total accuracy of each measurement as represented by
error bars in Fig.~\ref{fig:summary}.

\begin{table}[t]
\caption[Pull and weight of each measurement]{The pull and weight for each of the
  inputs, as obtained from the combination with the BLUE method to
  determine the average top quark mass.}
\begin{center}
\renewcommand{\arraystretch}{1.30}
{\tiny
\begin{tabular}{l|ccccc|cccccc|c}
\hline 
\hline
       & \multicolumn{5}{c}{{\RunI} published} 
       & \multicolumn{6}{|c|}{{\RunII} published} 
       & \multicolumn{1}{c}{{\RunII} preliminary}  \\
       & \multicolumn{3}{c}{ CDF } 
       & \multicolumn{2}{c}{ D\O\ }
       & \multicolumn{4}{|c}{ CDF } 
       & \multicolumn{2}{c|}{ D\O\ } 
       & \multicolumn{1}{c}{ CDF } \\
       &  $\ell$+jets    &    $\ell\ell$ &   alljets &    $\ell$+jets  &   $\ell\ell$   &   $\ell$+jets  &   $\ell\ell$   & alljets   &     Lxy   &  $\ell$+jets      & $\ell\ell$  &  MEt \\
\hline



Pull   & $+0.40$  & $-0.51$  & $+1.11$  & $+1.32$  & $-0.38$  & $-0.51$  & $-0.82$  & $-0.41$  & $-0.67$ & $1.42$ & $+0.30$   & $+0.45$  \\
Weight [\%]  & $-4.7$ & $-1.1$ & $-0.9$ & $+0.4$ & $-0.2$ & $+62.0$ & $-0.3$ & $+10.5$ & $+0.22$ & $+20.6$ & $+1.4$     & $+11.9$  \\
\hline \hline
\end{tabular}
}
\end{center}
\label{tab:stat} 
\end{table} 

\begin{figure}[p]
\begin{center}
\includegraphics[width=0.8\textwidth]{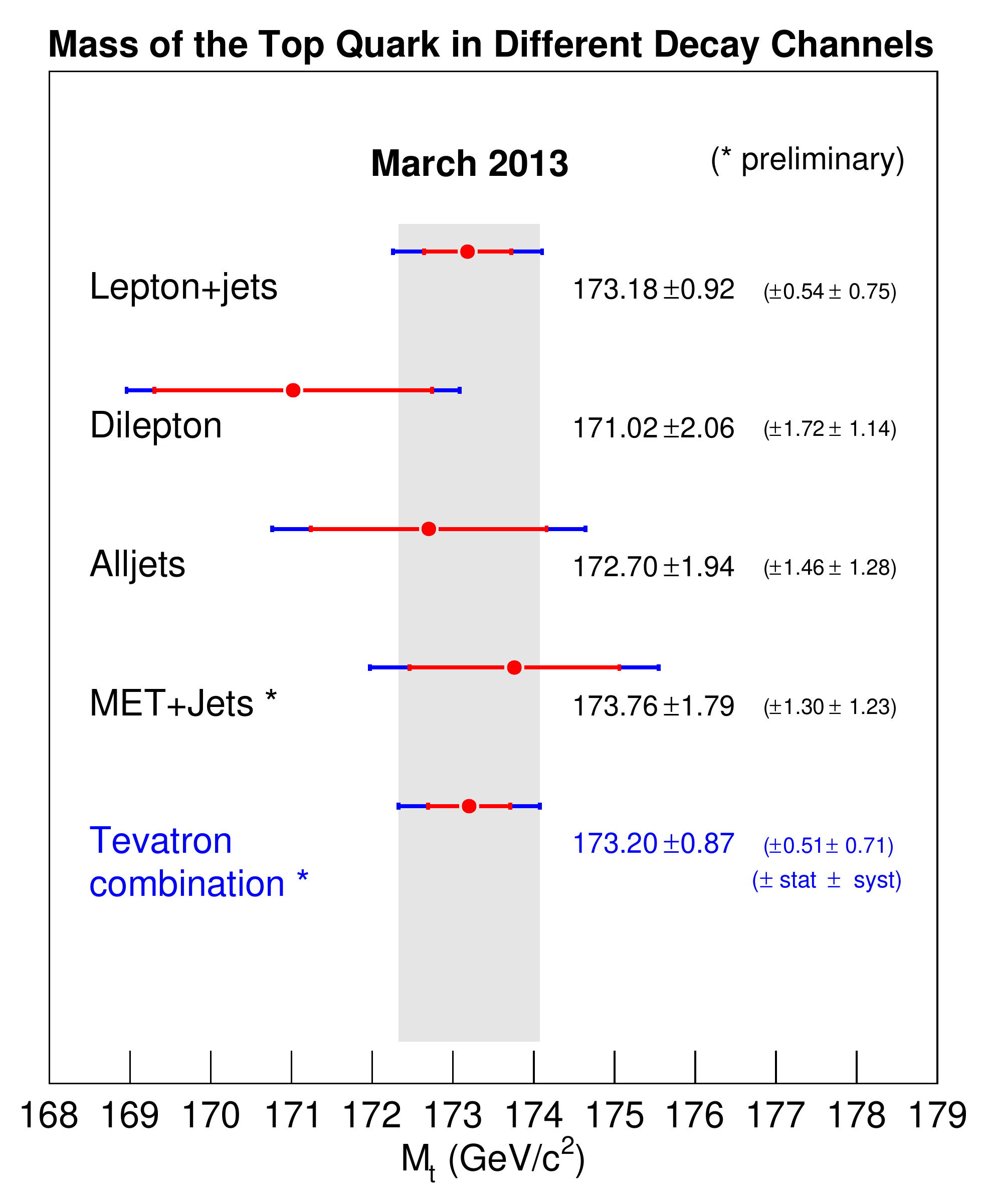}
\end{center}
\caption[Mtop in each channel]{Summary of the combination of the 12
top-quark measurements by CDF and D\O\ for different final states.}
\label{fig:three_observables} 
\end{figure}

No input has an anomalously large pull. 
It is, however, still
interesting to determine the top-quark mass separately 
in the alljets, $\ell$+jets, $\ell\ell$, and MEt channels (leaving out
the $\Lxy$ measurement).
We use the same methodology,
inputs, uncertainty categories, and correlations as described above, but fit 
the four physical observables, \MTjj, \MTlj, \MTll, and \MTmet\ separately.
The results of these combinations are shown in
Figure~\ref{fig:three_observables} and Table~\ref{tab:three_observables}.

Using the results of Table~\ref{tab:three_observables} 
 we calculate the following chi-squared values including correlations:
$\chi^{2}(\ell+{\rm jets}-\ell\ell)=1.30/1$, $\chi^{2}(\ell+{\rm
  jets}-{\rm alljets})=0.07/1$,  $\chi^{2}(\ell+{\rm jets}-{\rm MEt})=0.11/1$, 
$\chi^{2}(\ell\ell-{\rm alljets})=0.42/1$,
  $\chi^{2}(\ell\ell-{\rm MEt})=1.22/1$, and $\chi^{2}({\rm
  alljets}-{\rm MEt})=0.19/1$.  
These correspond to chi-squared probabilities 
of 25\%, 79\%, 74\%, 52\%, 27\%, and 66\%  respectively, indicating
that the top-quark mass determined in each decay channel is consistent
in all cases.

\begin{table}[t]
\caption[Mtop in each channel]{Summary of the combination of the 12
measurements by CDF and D\O\ in terms of four physical quantities,
the mass of the top quark in the alljets, $\ell$+jets,  $\ell\ell$, and MEt decay channels. }
\begin{center}
\renewcommand{\arraystretch}{1.30}
\begin{tabular}{ccrrrr}
\hline\hline

Parameter & Value (\GeVc2) & \multicolumn{4}{c}{Correlations} \\
               &                                 & $\MTjj$ &    $\MTlj$  &  $\MTll$ & $\MTmet$ \\ \hline
$\MTjj$ & $172.7\pm 1.9$    & 1.00       &                   &      & \\
$\MTlj$ & $173.2\pm 0.9$    & 0.25       &    1.00       &     &\\
$\MTll$ & $171.0\pm 2.1$    & 0.19        &    0.41      & 1.00 & \\
$\MTmet$& $173.8\pm 1.8$    & 0.13        &    0.26      & 0.18 & 1.00 \\
\hline\hline
\end{tabular}
\end{center}
\label{tab:three_observables}
\end{table}

To test the influence of the choices in modeling the
correlations, we performed a cross-check by changing all non-diagonal
correlation 
coefficients of the correlation matrix defined in Section~\ref{sec:corltns} from 100\% to 50\% and re-evaluated the combination. 
The result of this large variation of degree of correlation is a
$\gevcc{+0.19}$ shift of the top-quark mass and reduces the total
uncertainty negligibly.
The chosen approach is therefore conservative.


We also performed separate combinations of all the CDF and D\O\ measurements. The results of these combinations are
$\gevcc{172.72~\pm~0.93}$ for CDF and $\gevcc{174.89~\pm~1.42}$ for D\O. Taking all correlations into account, 
we calculate the chi-square value $\chi^{2}(CDF-D\O)=2.25/1$ corresponding to a probability of 13\%.

\section{Summary}
\label{sec:summary}

An update of the combination of measurements of the mass of the top quark
from the Tevatron experiments CDF and D\O\ has been presented.  This preliminary
combination includes five published \RunI\ measurements, six published
\RunII\ measurements, and  
one preliminary \RunII\ measurement, but the majority of these
measurements are not yet performed on the full datasets available.  Taking into
account the statistical and systematic uncertainties and their
correlations, the preliminary result for the Tevatron average is
  $\MT=\gevcc{\measStatSyst{173.20}{0.51}{0.71}}$,
where the total uncertainty is obtained assuming Gaussian systematic uncertainties.
The central value is 0.02\,GeV/$c^2$  higher than our July 2012
average~\cite{TeVTopComboPRD} of 
$\MT=173.18\pm0.94$\,GeV/$c^2$.
%
%
Adding in quadrature the statistical and systematic uncertainties
yields a total uncertainty of $\gevcc{0.87}$ which represents an
improvement of $8\%$.

The mass of the top quark is now known with a relative precision of
0.50\%, limited by the systematic uncertainties, which are dominated by
the jet energy scale uncertainty.  
This result will be further improved when all analysis channels from  
CDF and D\O\ using the full Run~II data set are finalized.
 

\section{Acknowledgments}
\label{sec:ack}

We thank the Fermilab staff and the technical staffs of the
participating institutions for their vital contributions. 
This work was supported by  
DOE and NSF (USA),
CONICET and UBACyT (Argentina), 
CNPq, FAPERJ, FAPESP and FUNDUNESP (Brazil),
CRC Program, CFI, NSERC and WestGrid Project (Canada),
CAS and CNSF (China),
Colciencias (Colombia),
MSMT and GACR (Czech Republic),
Academy of Finland (Finland),
CEA and CNRS/IN2P3 (France),
BMBF and DFG (Germany),
Ministry of Education, Culture, Sports, Science and Technology (Japan), 
World Class University Program, National Research Foundation (Korea),
KRF and KOSEF (Korea),
DAE and DST (India),
SFI (Ireland),
INFN (Italy),
CONACyT (Mexico),
NSC(Republic of China),
FASI, Rosatom and RFBR (Russia),
Slovak R\&D Agency (Slovakia), 
Ministerio de Ciencia e Innovaci\'{o}n, and Programa Consolider-Ingenio 2010 (Spain),
The Swedish Research Council (Sweden),
Swiss National Science Foundation (Switzerland), 
FOM (The Netherlands),
STFC and the Royal Society (UK),
and the A.P. Sloan Foundation (USA).

\clearpage

\providecommand{\href}[2]{#2}\begingroup\raggedright\endgroup



\end{document}